\documentclass[
    ,final            
  ]
  {aipproc}
  
\layoutstyle{8x11double}

\newcommand{\mHt}{\rm{H}_{2}}
\newcommand{\hd}{\rm{HD}}
\newcommand{\Hm}{\rm{H^{-}}}
\newcommand{\mH}{\rm{H}}
\newcommand{\Hp}{\rm{H^{+}}}
\newcommand{\Dp}{\rm{D^{+}}}

\begin{document}

\title{Chemistry and cooling in metal-free and metal-poor gas}


\author{Simon Glover}{
address={Astrophysikalisches Institut Potsdam, An der Sternwarte 16, 
D-14482 Potsdam, Germany},
}

\begin{abstract}
I summarize four of the most important areas of uncertainty in the study of the chemistry
and cooling of gas with zero or very low metallicity. These are: i) the importance and
effects of HD cooling in primordial gas; ii) the importance of metal-line and dust cooling
in low metallicity gas; iii) the impact of the large uncertainties that exist in the rate
coefficients of several key reactions involved in the formation of $\mHt$; and iv)
the effectiveness of grain surface chemistry at high redshifts.
\end{abstract}

\maketitle


\section{Introduction}
Astrochemistry -- chemistry in an astrophysical context -- is a huge subject, owing to
the extremely large number of reactions that are possible, and the complexity of the
resultant reaction networks. For instance, the reaction sets used to 
study the chemistry of local molecular clouds typically include more than 300 
reactants and more than 4000 reactions simply to represent
the gas-phase chemistry. The inclusion of grain surface chemistry makes things
even more complex. However, when studying the formation
of the earliest generations of stars, our primary concern is to properly model the
thermal behaviour of the gas, as this strongly influences its hydrodynamical evolution.
We can therefore focus our attention solely on the chemistry of the major coolants, 
allowing us to sidestep much of the chemical complexity, and to
include only those reactants and reactions in our model that
are necessary for understanding the thermal evolution of the gas. 

An influential study of the chemistry of metal-free gas performed along these lines
by \cite{aanz97} demonstrated that out of the many hundreds of possible reactions,
only 28 were needed to accurately model the thermal evolution of the gas
over a wide range of parameter space. Subsequently, other authors have added
a few additional reactions to this simplified model that are very important in some
circumstances (e.g.\  HD formation and destruction, \cite{nu02}; three-body $\mHt$
formation, \cite{abn02}), but not in others.

A similar approach has recently been applied to metal-poor gas by \cite{gj07}
and \cite{g07a}, under the assumption that the dominant coolants in metal-poor
gas are the same as those in the local atomic and molecular ISM (i.e.\ carbon,
oxygen and silicon fine-structure lines, CO and ${\rm H_{2}O}$ rotational and
vibrational emission, and dust). In this case, the required number of reactions
is an order of magnitude larger than in the metal-free case, but is still
small enough to be computationally tractable.

Nevertheless, despite our success at identifying the most important reactions
and hence simplifying the treatment of the chemistry to the point where it can
be modelled self-consistently within large, three-dimensional hydrodynamical
simulations, a number of issues remain unresolved. In this contribution, I
briefly discuss a few of what I consider to be the most important areas of
ongoing uncertainty or confusion in the study of the chemistry and cooling 
of metal-free and metal-poor gas. 

\section{Problem 1: the role of HD cooling}
At temperatures $200 < T < 10000 \: {\rm K}$ in a metal-free gas (or in low-density
metal-poor gas; \cite{jgkm07}), molecular hydrogen is the most abundant molecular
species and is also the dominant coolant. However, at low temperatures, $\mHt$
cooling becomes increasingly ineffective, owing to the large energy difference
that exists between its $J=0$ and $J=2$ rotational levels, and the fact that the lower
energy $J=1 \rightarrow 0$ transition is strongly forbidden. At these low temperatures,
HD cooling is far more effective, in the sense that the cooling rate per molecule of
HD is much larger than the cooling rate per molecule of $\mHt$. However, the small
size of the cosmic D/H ratio means that in general HD is far less abundant than $\mHt$.
Because of this, for a long time it was unclear whether or not HD was ever an important
coolant in practice. In the past few years, however, considerable effort has been devoted
to exploring the consequences of HD cooling in the early Universe, and its role has now
become much clearer. 

In \cite{bcl02}, the impact of HD cooling in the first star-forming protogalaxies was 
examined, and it was shown that given a D/H ratio consistent with observationally 
measured values, HD cooling was unimportant. More recently, the role of HD cooling within 
a wider range of protogalaxies was examined (\cite{ripa07}), with the conclusion that
in very low mass protogalaxies ($M < 3 \times 10^{5} \: {\rm M_{\odot}}$), HD cooling may 
be marginally important. 

These calculations typically adopted an initial temperature and chemical composition 
for the gas consistent with the values applicable to the undisturbed IGM. However, 
other studies that start with initially ionized gas (e.g.\ \cite{jb06,yokh07}) find 
that in this case, HD cooling {\em is} important, and that it can often cool the gas 
all the way down to the temperature floor set by the CMB. 

\begin{figure}
\includegraphics[angle=270,totalheight=.3\textheight]{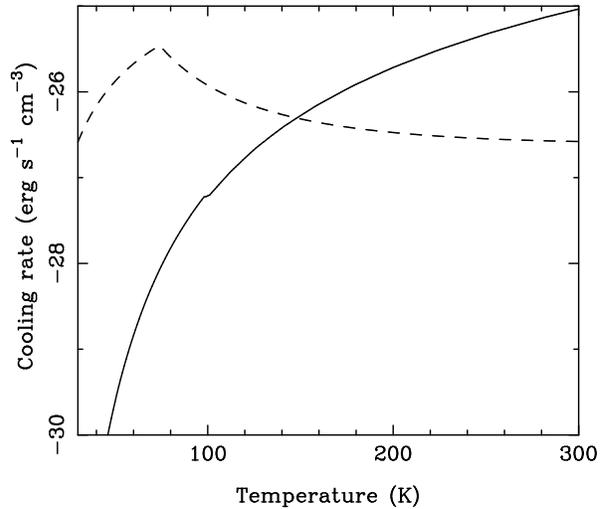}
\caption{Cooling rates of $\mHt$ (solid line) and $\hd$ (dashed line)
in a gas with atomic hydrogen number density $n_{\mH} = 100 \: 
{\rm cm^{-3}}$. The fractional $\mHt$ abundance is fixed at 
$x(\mHt) = 10^{-3}$, while the HD/$\mHt$ ratio, $x(\hd) / x(\mHt)$, 
is calculated assuming that reactions~\ref{d_react_1} 
and \ref{d_react_2} have reached equilibrium. The maximum
abundance of HD is constrained to be no larger than the total
deuterium abundance, here taken to be $x(\rm D, tot) = 2.6 
\times 10^{-5}$. This constraint is responsible
for the sudden downturn in the HD cooling rate at low $T$.
\label{hdcool}}
\end{figure}

Why do these two sets of calculations differ so much on the importance of HD?
The answer lies in the low temperature chemistry of the HD molecule. In most
circumstances, its formation and destruction are dominated by the ion-neutral
exchange reactions:
\begin{eqnarray}
\mHt + \Dp & \rightarrow & \hd + \Hp,  \label{d_react_1} \\
\hd + \Hp & \rightarrow & \mHt + \Dp. \label{d_react_2}
\end{eqnarray}
Reaction~\ref{d_react_1} is exothermic, but reaction~\ref{d_react_2}
is endothermic by 462~K (\cite{gp02}). As a result, at low gas temperatures, 
considerable chemical fractionation occurs: the equilibrium value of the 
HD/$\mHt$ ratio is enhanced over the cosmic D/H ratio by a factor 
$\sim e^{462 / T}$. The effect of this on the cooling of the gas is illustrated in 
Figure~\ref{hdcool}, where I plot the cooling rate of the gas as a function
of temperature for gas with $n = 100 \: {\rm cm^{-3}}$ and $x(\mHt) = 10^{-3}$,
and where $x(\hd) / x(\mHt)$ is calculated assuming that reactions~\ref{d_react_1}
and \ref{d_react_2} have reached equilibrium. In practice, fractionation may not
proceed fully to equilibrium owing to the depletion of the necessary $\Dp$ ions
from the gas by recombination, so this plot should be regarded as giving an
upper limit on the effectiveness of $\hd$ cooling at this density.
Figure~\ref{hdcool} demonstrates that at $T \sim 150 \: {\rm K}$, the
increase in the HD abundance due to fractionation more than offsets the
decrease in the $\mHt$ and $\hd$ cooling rates per molecule. 
Consequently, gas that can cool to this critical temperature will begin
to cool even more effectively, and thus can easily reach temperatures
as low as $T_{\rm CMB}$. The critical temperature that must be reached
for HD cooling to take over in this fashion varies with $n$, but is never
larger than 250~K, and exceeds 200~K only for gas with 
$n > 10^{4} \: {\rm cm^{-3}}$.

It is now straightforward to explain why HD cooling becomes important
in gas cooling from an initially highly ionized state, but not for gas cooling
from a low-ionization state. In the former case, it has long been recognised
that the out-of-equilibrium nature of hydrogen recombination in the cooling
gas leads to enhanced production of $\mHt$. As a result,
gas cooling from a high-ionization state can reach a lower temperature than
gas cooling from a low-ionization state. Although the temperature difference
is not large, it turns out that in the former case, the gas can usually
cool to $T_{\rm crit}$, at which point HD takes over, while in the latter case
it falls somewhat short, typically never cooling below $T \sim 200 \: {\rm K}$.

This explanation accounts for the differences in behaviour observed in 
previous numerical simulations with different initial conditions, but also
prompts a couple of further questions. First, what effect does the presence of
an ultraviolet background in the Lyman-Werner bands of $\mHt$ have
on the ability of the gas to cool below $T_{\rm crit}$? Most previous
simulations that include the effects of HD cooling have assumed that 
the ultraviolet background is zero, or at least is too small to be significant.
Recently, however, the impact of a Lyman-Werner background on the
effectiveness of HD cooling was investigated in \cite{yoh07}
using a one-zone chemical model. This investigation suggests that if the 
field strength is greater than $3 \times 10^{-22} \: {\rm erg} \: {\rm
s^{-1}} \: {\rm cm^{-2}} \: {\rm Hz^{-1}} \: {\rm sr^{-1}}$, then the gas
never cools below 200~K, and so HD cooling never becomes 
important. However, this calculationsuffers from the problem -- common 
to all one-zone models -- that the dynamical evolution is not solved
for self-consistently with the thermal evolution. Further investigation of
this issue using a proper hydrodynamical approach would be extremely
useful.

Second, does the extra cooling at low temperatures provided by the HD 
actually promote fragmentation? Judging by recent work (e.g.\ \cite{yokh07}),
the answer is no: the additional cooling reduces the fragment mass scale, 
but despite this only a single fragment forms in each halo. However, the 
sharp rise in the cooling rate at $T < T_{\rm crit}$ apparent in 
Figure~\ref{hdcool} suggests that HD cooling may render the gas 
thermally unstable in some circumstances, which would tend to promote 
fragmentation.

\section{Problem 2: the importance of metals}
One of the most important unsolved problems in the study of star formation in
very low metallicity gas is the role that cooling from gas-phase metals and
dust grains plays
in determining the stellar IMF. A popular hypothesis (introduced in \cite{bfcl01})
holds that it is the enrichment of the star-forming gas with metals to a level
greater than a critical metallicity ${\rm Z_{crit}}$ that brings about a transformation
in the IMF from one dominated by massive stars to one more closely resembling
the standard Salpeter IMF. The causal link between metal enrichment and the
formation of lower mass stars is the greater effectiveness of metals as coolants
(particularly at low temperature) compared to $\mHt$, which is supposed to 
enable the gas to gravitationally fragment down to much smaller mass scales.

Two different forms of metal-related cooling have attracted considerable
attention: fine-structure cooling from C and O (and in some
cases also Si and Fe; \cite{ss06}),
which is effective primarily at densities $n < 10^{6} \: {\rm cm^{-3}}$
\cite{bfcl01,ss06,bl03,ss07}, and dust cooling, which is effective at much
higher densities \cite{sch02,om05,to06,cgk07}.

The effectiveness of fine-structure cooling at promoting fragmentation
remains highly uncertain. The original simulations of \cite{bfcl01} found
fragmentation at ${\rm Z} = 10^{-3} \: {\rm Z_{\odot}}$ but not at 
${\rm Z} = 10^{-4} \: {\rm Z_{\odot}}$, implying that $10^{-4} < {\rm Z_{crit}}
< 10^{-3} \: {\rm Z_{\odot}}$. However, these simulations did not include
the effects of $\mHt$ cooling. Very similar simulations have recently been
performed that do include $\mHt$ cooling (\cite{jkgm07}). In this case, 
fragmentation occurs even in runs with ${\rm Z} = 0$, suggesting that the
fragmentation seen in this type of simulation owes more to the choice of 
initial conditions than to metal-enrichment. Furthermore, simulations
performed using alternative initial conditions in which the gas is hot
and ionized (\cite{jkgm07}) find no fragmentation at 
${\rm Z} \leq 10^{-3} \: {\rm Z_{\odot}}$, whereas simulations by 
\cite{ss07}, performed using yet another set of initial conditions, 
\emph{do} find fragmentation at ${\rm Z} = 10^{-3} \: {\rm Z_{\odot}}$. 
The logical conclusion to be drawn
from these contradictory results is that whether or not metal-enriched gas
fragments depends to a large extent on its initial state, and that without
a better understanding of this, one cannot be certain about the importance
of fine-structure cooling in driving fragmentation. 

In the case of dust cooling, recent hydrodynamical simulations (described
in \cite{to06}, \cite{cgk07} and elsewhere in these proceedings) find that 
the gas readily fragments following the onset of dust cooling, provided that
the available dust surface area per unit gas mass exceeds one part in 
$10^{5}$ of the value in solar metallicity gas. Although the simulations
of \cite{to06}, which consider the special case of gas with zero angular
momentum, find only limited fragmentation, the simulations of \cite{cgk07},
which consider the more general case of non-zero angular momentum,
find a high level of fragmentation, with $\sim 100$ distinct fragments 
formed within only a few hundred years of the formation of the first object.

Nevertheless, despite these recent results, a number of serious questions 
remain. To begin with, both sets of simulations were performed using a
tabulated equation of state, based on the thermal 
evolution of the gas found in one-zone models of low-metallicity protostellar
collapse (e.g.\ \cite{om05}). However, one-zone models of this type do
not self-consistently couple the thermal and dynamical evolution of the
gas, and also cannot properly account for effects such as rotation.
Conclusions based on one-zone models of collapse in metal-free gas
have in the past proved to be highly misleading (see e.g.\ the discussion
in \cite{g05}). Moreover, by using a tabulated equation of state, we 
implicitly assume that the gas temperature adjusts instantaneously 
to any changes in density; in practice, such changes are not instantaneous,
but instead take place over a thermal timescale. It is possible that by
making this implicit assumption, we are overestimating the propensity of
the gas to fragment. Work aimed at clarifying these issues by solving the
full thermal energy equation self-consistently within the hydrodynamical
simulations is currently ongoing.

The other major concern about the current results is that they do not
take into account feedback in the form of radiation from the accretion
shocks surrounding the newly formed protostars. In local star-forming
regions, this form of feedback may play a crucial role in stabilizing 
massive cores against sub-fragmentation (\cite{krum06}), and so its
effects should clearly also be included in the high redshift simulations.
However, there are reasons to believe that its effects will be far less
significant in a high-redshift, low-metallicity context. For one thing, 
the gas in these low-metallicity cores that cools rapidly and fragments 
is typically not optically thick, and so will absorb only a fraction of 
the radiation emitted by the accreting protostars. For another, the
temperature of the dust in these objects is very much higher than in
local star-forming regions ($T_{\rm d} \sim 200 \: {\rm K}$ at the
relevant densities, compared to $T_{\rm d} \sim 10 \: {\rm K}$ locally),
and so a much higher energy input is required to significantly alter
the temperature of the gas. For these reasons, it seems unlikely that
the inclusion of feedback will significantly alter the outcome of the
simulations, but this expectation must nevertheless still be properly 
tested.

\section{Problem 3: rate coefficient uncertainties}
Our ability to construct accurate models of the chemical evolution of metal-free or
metal-poor gas is constrained by the level of accuracy to which the rate coefficients
of the key chemical reactions have been determined. Most of the important reactions 
involved in the chemistry of $\mHt$  and HD have well-determined rate coefficients
(\cite{aanz97}, \cite{gp98}). However, there are a few important exceptions. 

Two of these exceptions are the associative detachment of $\Hm$ with H
\begin{equation}
\Hm + \mH \rightarrow \mHt + {\rm e^{-}}, \label{ad_react}
\end{equation}
and the mutual neutralization of $\Hm$ with $\Hp$
\begin{equation}
\Hm + \Hp \rightarrow \mH + \mH. \label{mn_react}
\end{equation}
The available experimental and theoretical data on the low temperature 
($T < 10^{4} \: {\rm K}$) behaviour of these two reactions is summarized in
\cite{gsj06}. In both cases, the rate coefficients are uncertain by at least 
a factor of a few. In gas with a low fractional ionization, this 
uncertainty is unimportant, as reaction~\ref{ad_react} occurs at a much faster rate than
reaction~\ref{mn_react}, regardless of which of the possible values are selected 
for the two rate coefficients. In gas recombining from an initially ionized state,
on the other hand, the uncertainties in reactions~\ref{ad_react} and \ref{mn_react}
combine to render the $\mHt$ abundance uncertain by as much as an order
of magnitude at some points during the chemical evolution of the gas. Some 
possible consequences of this were also explored in \cite{gsj06}.

\begin{figure}
\includegraphics[angle=270,totalheight=.3\textheight]{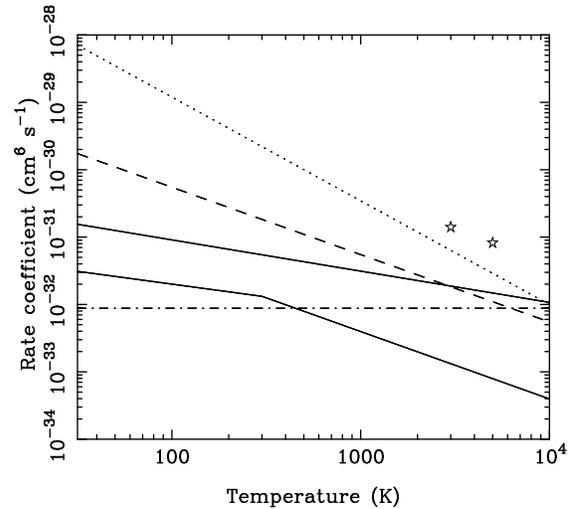}
 \caption{Three-body $\mHt$ formation rate coefficients, plotted as
 a function of temperature. Key: A -- lower solid line; B -- dashed
 line; C -- dot-dashed line; D -- dotted line; E -- star symbols;
 F -- upper solid line. \label{3body}}
\end{figure}

Another example is the large uncertainty that exists in the rate coefficient of the
three-body $\mHt$ formation reaction
\begin{equation}
\mH + \mH + \mH \rightarrow \mHt + \mH, \label{3b_react}
\end{equation}
the dominant source of $\mHt$ in gas at densities $n > 10^{8} \: {\rm cm^{-3}}$. In
Figure~\ref{3body}, I show a selection of rate coefficients quoted in the literature for 
this reaction. The corresponding numerical values and references are listed in
Table~\ref{tab:3body}.  This compilation includes one new rate coefficient (F), 
presented here for the first time. This was computed using the same detailed balance approach
as in \cite{fh07}, but illustrates the effect of using an $\mHt$ collisional dissociation 
rate taken from \cite{msm96} in place of the rate from \cite{jgc67} used in \cite{fh07}.

\begin{table}
\begin{tabular}{llc}
\hline
   \tablehead{1}{r}{b}{ID}
& \tablehead{1}{r}{b}{Rate coefficient
(${\rm cm^{6}} \: {\rm s^{-1}}$)}
& \tablehead{1}{r}{b}{Reference} \\
\hline
A & $1.14 \times 10^{-31} T^{-0.38}$ \hspace{20pt} $T \leq 300 \: {\rm K}$ & \cite{abn02} \\
 & $3.9 \times 10^{-30} T^{-1.0}$ \hspace{28pt} $T > 300 \: {\rm K}$ & \\
B & $5.5 \times 10^{-29} T^{-1.0}$ & \cite{pss83} \\ 
C & $8.8 \times 10^{-33}$ & \cite{cw83} \\
D & $1.44 \times 10^{-26} T^{-1.54}$ & \cite{fh07} \\
E & $1.4 \times 10^{-31}$ \hspace{50pt} $T = 3000 \: {\rm K}$ & \cite{schw90} \\
 & $8.2 \times 10^{-32}$ \hspace{50pt} $T = 5000 \: {\rm K}$ & \\
F & $7.7 \times 10^{-31} T^{-0.464}$ & This work \\
\hline
\end{tabular}
\caption{Rate coefficients for three-body $\mHt$ formation}
\label{tab:3body}
\end{table}

It is clear from Figure~\ref{3body} that although there is reasonable agreement
between many of the rates at 5000~K, there is a substantial uncertainty in the
rate in the temperature range $200 < T < 2000 \: {\rm K}$ relevant for population
III star formation. Moreover, there is no sign that the rates are converging: two
of the most recent determinations (A and D) are the two most widely discrepant
rates. 

The $n^{3}$ density dependence of the three-body $\mHt$ formation process
ensures that in collapsing primordial protostellar cores, the gas will rapidly
become fully molecular at high densities, regardless of the rate coefficient
adopted. However, the choice of rate coefficient will strongly influence 
the three-body $\mHt$ formation heating rate, which is a major energy input
into the gas at these densities, and may therefore also affect the further 
dynamical evolution of the gas. Simulations aimed at exploring the 
consequences of this uncertainty are currently in progress. 

\section{Problem 4: grain surface chemistry}
The key role that dust grains play in the chemistry of molecular gas in the
Milky Way and in other local galaxies has been understood for more
than 40 years. Very few of the $\mHt$ molecules in these systems were 
formed in the gas phase via the $\Hm$ or ${\rm H_{2}^{+}}$ pathways that are so 
important in primordial gas. Instead, most were formed by reactions between
hydrogen atoms adsorbed on the surface of interstellar dust grains. Moreover,
$\mHt$ is far from the only molecule that can form in this manner.

However, the importance of grain surface chemistry at high redshift remains an open
question. There are two key issues. The first is that the amount of
dust present in high redshift, low metallicity systems is not well constrained.
A number of calculations have been performed that explore the formation of 
dust grains in high-redshift supernova remnants produced by metal-free
progenitors (\cite{tf01,noz03,sch04,bs07,noz07}), but the results 
of these calculations do not agree on the details of the resulting grain
size distribution, and so the available grain surface area per unit gas mass
remains uncertain.

The second key issue is the efficiency of molecule formation on high 
redshift dust grains. On local dust grains, most chemistry occurs through
the interaction of van der Waals bonded (or `physisorbed') atoms. The
evaporation timescale for these atoms is typically orders 
of magnitude longer than the accretion timescale, and so the atoms 
have plenty of time in which to react. At high $z$, however, the dust 
grains are warmer, owing to the effects of heating by the CMB, 
and so the evaporation timescales, which depend exponentially on
$T_{\rm dust}$, are vastly smaller. For important atoms such as
H, C or O, this means that $t_{\rm evap} \ll t_{\rm acc}$, and so the
likelihood of there being more than a single atom physisorbed
on a given grain at any given time is extremely small. Taken at face 
value, this implies that the efficiency of molecule formation 
on high redshift dust grains should be extremely small. 

If one applies a similar argument 
to warm dust in local photodissociation regions, then one predicts
a similar outcome: molecule formation, and in particular the formation
of $\mHt$, should be extremely inefficient there. However, this 
conclusion conflicts with observationally-derived constraints on the
$\mHt$ formation rate in these regions, suggesting that some or
all of the $\mHt$ formed in these regions is formed from H atoms
that are chemically bonded (or `chemisorbed') to the surface of the
grains (\cite{hab04}).  A detailed model for the formation of $\mHt$
on grains that have both physisorption and chemisorption binding
sites has recently been constructed by \cite{ct04} and applied to
the question of high-redshift $\mHt$ formation by \cite{cs04}. The
results of this model suggest that once chemisorption is taken into
account, $\mHt$ formation remains efficient even at very high redshifts.
However, this model is not without its critics (e.g.\ \cite{vid05}), and as 
yet no similar modelling effort has been completed for the formation 
of any species other than $\mHt$, although work along these lines is 
currently under way.


\begin{theacknowledgments}
I would like to thank Tom Abel, Daniel Savin and Naoki Yoshida for many
interesting and enjoyable conversations about the chemistry of metal-free
gas. 
\end{theacknowledgments}


\end{document}